\documentstyle[12pt,graphicx,subfigure]{article}
\def\Journal#1#2#3#4{{#1} {\bf #2}, #3 (#4)}

\def\NPB{{\em Nucl. Phys.} B}
\def\PLB{{\em Phys. Lett.}  B}
\def\PRL{\em Phys. Rev. Lett.}
\def\PRD{{\em Phys. Rev.} D}

\def\beq{\begin{equation}}
\def\eeq{\end{equation}}

\def\gmin2{(g-2)_\mu}

\catcode`\@=11 

\def\lsim{\mathrel{\mathpalette\@versim<}}
\def\gsim{\mathrel{\mathpalette\@versim>}}
\def\@versim#1#2{\vcenter{\offinterlineskip
    \ialign{$\m@th#1\hfil##\hfil$\crcr#2\crcr\sim\crcr } }}

\def\PRL{Phys. Rev. Lett.}

\def\beq{\begin{equation}}
\def\eeq{\end{equation}}
\def\beqn{\begin{eqnarray}}
\def\eeqn{\end{eqnarray}}
\def\Journal#1#2#3#4{{#1} {\bf #2}, #3 (#4)}

\def\NPB{{\em Nucl. Phys.} B}
\def\PLB{{\em Phys. Lett.}  B}
\def\PRL{\em Phys. Rev. Lett.}
\def\PRD{{\em Phys. Rev.} D}

\def\lsim{\ ^<\llap{$_\sim$}\ }
\def\gsim{\ ^>\llap{$_\sim$}\ }
\def\r2{\sqrt 2}
\def\beq{\begin{equation}}
\def\eeq{\end{equation}}
\def\beqn{\begin{eqnarray}}
\def\eeqn{\end{eqnarray}}

\def\sinW2{\sin^2\theta_W}

\def\mz2{M_{z}^2}
\def\c2b{\cos 2\beta}

\def\mz{M_z}

\def\Fq2{F_{2}(q^2)}
\def\beq{\begin{equation}}
\def\eeq{\end{equation}}

\def\gmin2{(g-2)_\mu}

\def\sec2w{sec^2\theta_W}

\def\lsim{\ ^<\llap{$_\sim$}\ }
\def\gsim{\ ^>\llap{$_\sim$}\ }
\def\r2{\sqrt 2}
\def\beq{\begin{equation}}
\def\eeq{\end{equation}}
\def\beqn{\begin{eqnarray}}
\def\eeqn{\end{eqnarray}}

\def\sinW2{\sin^2\theta_W}

\def\mz2{M_{z}^2}
\def\c2b{\cos 2\beta}

\def\mz{M_z}

\def\Fq2{F_{2}(q^2)}
\def\sq2{\sqrt{2}}

\def\sec2w{sec^2\theta_W}
\begin{document}

\begin{titlepage}

\begin{center}
{\large {~\bf Corrections to b, t quark masses and $\tau$ lepton mass in SUGRA 
including CP phases}}\\
\vskip 0.5 true cm
\vspace{2cm}
\renewcommand{\thefootnote}
{\fnsymbol{footnote}}
 Tarek Ibrahim$^{a,b}$ and Pran Nath$^{b}$  
\vskip 0.5 true cm
\end{center}

\noindent
{a. Department of  Physics, Faculty of Science,
University of Alexandria,}\\
{ Alexandria, Egypt\footnote{: Permanent address }}\\ 
{b. Department of Physics, Northeastern University,
Boston, MA 02115-5000, USA} \\
\vskip 1.0 true cm
\centerline{~ Abstract}
\medskip
A brief review is given of recent analyses of the effects of CP phases
on the supersymmetric QCD and supersymmetric electroweak contibutions 
to the $b$ and $t$ quark masses and to the $\tau$ lepton mass in 
SUGRA models.
 The effects of CP phases on the supersymmetric contributions are
 found to be significanlty large for the b quark mass as they can change
 both its sign and its magnitude. Thus 
 with the inclusion of CP phases the supersymmetric correction to
 the b quark mass can be as much as fifty percent or more of the 
 total b quark mass. For
 the case of the $\tau$ lepton, the effects of CP phases on the
 supersymmetric correction is also relatively large as it can again
 affect both the sign and the magnitude of the $\tau$ mass 
 correction. However,
 in this  case the overall correction is found to be only a few 
 percent. The effect of CP phases on SUSY contribution to the $t$ 
 quark mass was also investigated. However, in this case the
 overall correction is less than a percent with or  without the
 inclusion of phases. These results have important implications
 for $b-\tau$ unification and for $b-t-\tau$ unification in the
 context of unified theories. 
 \end{titlepage}

\section{Introduction}
Supersymmetric corrections to the $b$, $t$ quark masses and  to the
 $\tau$ lepton mass are of great importance. These corrections
affect importantly analyses of $b-\tau$ and $b-t-\tau$ couplings
in unified models of particle interactions\cite{arason,baer}.
 Further, such corrections
can also affect a variety of low energy phenomena such as decays
of the Higgs into $b\bar b$, $\tau\bar\tau$, $c\bar c$ etc which
are potentials sources as signals of supersymmetry. 
In this talk we discuss the effect of CP phases on supersymmetric
corrections to the $b$, $t$ quark masses and to the $\tau$ lepton mass.
Previous analyses have not fully taken account of  the 
phases\cite{Hall:1993gn,carena94,Pierce:1996zz,carena2000,carena2002}
For the $b$ and $t$ quark masses these arise from the SUSY QCD and
SUSY electroweak contributions from the exchange of the gluino,
charginos and neutralinos. For the $\tau$ lepton mass they arise 
from the  SUSY electroweak contributions from the exchange of the 
charginos and the neutralinos. The CP phases in SUSY have a long
history. It was realized early on that the SUSY CP phases could
pose a severe problem in that they could generate large  contributions
to the electric dipole moments of the neutron and of the electron 
which may exceed the experimental limits\cite{nedm,eedm}. 
Additionally one now also has very stringent limits on the atomic
edms specifically on the edm of $Hg^{199}$ (see Ref.\cite{atomic}). 
Initially the technique
followed was to suppress the edm contributions by simply adjusting
the CP phases to be small\cite{ellis}. 
However, since then other ways have been
devised which allow one to suppress the  edms while allowing for
large phases\cite{na,bdm2,incancel,inbrane,chang,olive,inhg199}. 
One technique of interest here is the cancellation
mechanism\cite{incancel} wherein the edm contributions are suppressed by 
cancellations
among various contributions. With this mechanism one can allow for
large phases consistent with the current edm constraints. 
The process of cancellation is facilitated  by the presence of  
several phases. Thus the minimal supergravity model 
(mSUGRA)\cite{msugra} can
allow for two phases so that the parameter space including phases is
described by $m_0$, $m_{\frac{1}{2}}$, $|A_0|$, $\tan\beta$, 
$\alpha_{A_0}$ and $\theta_{\mu}$, where $m_0$ is the universal scalar
mass,  $m_{\frac{1}{2}}$ is the universal gaugino mass, $|A_0|$
is the universal trilinear coupling, $\tan\beta ={H_2}/{H_1}$
where $H_2$  gives mass to the up quark and $H_1$ gives mass to
the down quark and the lepton, $\alpha_{A_0}$ is the phase of $A_0$ and
$\theta_{\mu}$ is the phase of the Higgs mixing parameter $\mu$. 
However, in the presence of nonuniversalities the more general 
SUGRA model can accommodate more phases. Thus, for example,
due to a  non flat gauge kinetic energy function one may have
the $SU(3)_C\times SU(2)_L\times U(1)$ gaugino masses  which are
nonuniversal so that  
$\tilde m_i = | \tilde m_i| e^{i\xi_i} (i=1,2,3)$
In this case one  has  a greater
parameter space for the cancellation mechanism to operate.
The  presence of phases has typically a large effect on low
energy phenomenology and a small sample of these is given in 
Refs.~
\cite{pilaftsis,inhiggs,Carena:2001fw,kane,barger,zerwas,ing2,masiero1,dedes}
while a sample  of some more recent works are given in Ref.\cite{recent}
A more complete list of works can be seen in Ref.~\cite{insusy02}.

 We discuss now some details of the technique for the computation 
 of the SUSY mass correction to the quark and lepton masses.
We follow closely the analysis of Ref.\cite{inbtau}.
 We begin by noting that the pole mass $M_b$ which is the 
  physical mass of the b quark is related to the running mass 
  $m_b(M_b)$ as follows 
 
\beq 
M_b=(1+\frac{4\alpha_3(M_b)}{3\pi}+12.4\frac{\alpha_3(M_b)^2}{\pi^2})
m_b(M_b)
\eeq
where the corrections in the brace on the right hand side are the 
QCD corrections up to two loop level\cite{arason}.  The quantity
 $m_b(M_b)$ is obtained  from $m_b(M_Z)$ by use of renormalization
  group evolution. In the analysis here  we will focus on 
  the computation of $m_b(M_Z)$  which is the running
$b$ quark mass at the $Z$ scale. 
This quantity can be written in the
form 
\beq 
m_b(M_Z)=h_b(M_Z)\frac{v}{\sqrt 2}\cos\beta(1+\Delta_b) 
\eeq
where $\Delta_b$ is loop correction to $m_b$ 
\cite{Hall:1993gn,carena94,Pierce:1996zz,carena2002}.
We give now further details of our analysis.
  Our procedure is similar to that of 
 Ref\cite{carena2002}. We begin by noting that at the tree level
 one has a coupling of the b quark only to $H_1^0$. At the
 loop level the coupling of $H_1^0$ is modified and there is in 
 addition a correction to the couplings from $H_2^0$. Thus in the
 presence of loop corrections one can write the $b$ quark couplings
 as follows\cite{carena2002} 
\beqn
-L_{bbH^0}= (h_b+\delta h_b) \bar b_R b_L H_1^0 + 
\Delta h_b \bar b_R b_L H_2^0 + H.c.
\label{bbh}
\eeqn
Eq.~(\ref{bbh}) is the effective coupling of the $b$ quark to
the Higgs which can be used to compute the loop correction to 
$b$ quark mass. One finds 
\beqn
\Delta_b= [\frac{Re (\Delta h_b)}{h_b} \tan\beta 
+\frac{Re \delta h_b}{h_b} ]
\eeqn
Because of the presence of phases the loop corrections
$\Delta h_b$ and $\delta h_b$ will in general
be complex and the effective mass term for the $b$ quark will have
the form
  \beq	
  -L_{b}=(m_b^0+\Delta_1)\bar b b +i\Delta_2 \bar b \gamma_s b
  \label{complexbmass}
  \eeq
where $\gamma_5^{\dagger}=\gamma_5$ and $\Delta_{1,2}$ are real.
In Eq.~(\ref{complexbmass}) the $i\gamma_5$ term can be removed by  the 
 transformation $b=e^{i\frac{\theta}{2}\gamma_5}b'$ with an 
 appropriate choice of $\theta$. In the terms of the redefined
 fields one has
  \beq	
  -L_{b}= m_b \bar b' b',~~ 
  m_b=m_b^0+ \Delta_1+ \frac{1}{2} \frac{\Delta_2^2}{m_b^0} + ..
  \label{bmassexpansion}
  \eeq
 Eq.~(\ref{bmassexpansion}) shows that the $\Delta_2$ term 
 is essentially a higher order correction and can be safely ignored
 at the one loop level. 
 Similar considerations hold in the computation of loop corrections
 to the $\tau$ lepton mass and to the $t$ quark mass.
  
\section{CP Phase dependent loop corrections to the $b$, $t$ quark
masses and $\tau$ lepton mass}
\begin{figure}
\hspace*{-0.6in}
\centering
\includegraphics[width=9cm,height=4cm]{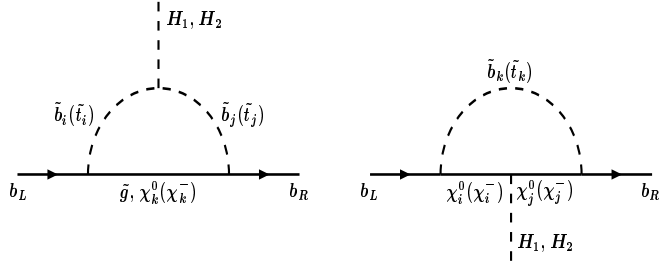}
\caption{One loop SUSY QCD and SUSY electroweak contribution to the 
b quark mass arising from  
exchange of gluino, charginos and neutralinos in the loop.}
\label{bloop}
\end{figure}

\begin{figure}
\hspace*{-0.6in}
\centering
\includegraphics[width=9cm,height=4cm]{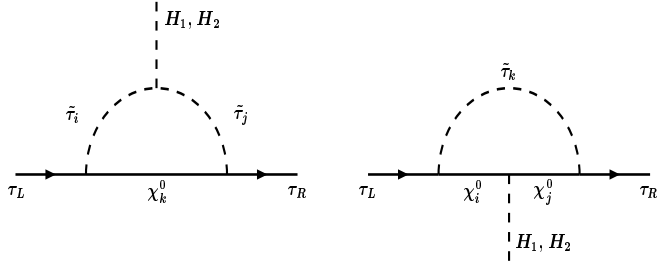}
\caption{One loop SUSY electroweak correction to the 
$\tau$ lepton mass arising from the  
exchange of charginos in the loop.}
\label{tauloop}
\end{figure}

We begin with a discussion of correction to the b quark mass.
Loop contributions to the b quark mass arise from the 
exchange of the gluino, chargino and neutralinos (see Fig.~\ref{bloop} )
\beqn 
\Delta_b = \Delta_b^{\tilde g} + \Delta_b^{\tilde \chi^+}
+ \Delta_b^{\tilde \chi^0} 
\eeqn
Here $\Delta_b^{\tilde g}$ is the gluino 
exchange contribution, $\Delta_b^{\tilde \chi^+}$ is  the
chargino exchange contribution, and $\Delta _b^{\tilde \chi^0}$ is 
the neutralino exchange contribution. The full expressions for
all these three contributions can be found in Ref.~\cite{inbtau}.
We will discuss here for
illustrative purposes the gluino exchange contribution explicitly.
The total correction to the b quark mass from gluino exchange 
including the effects of CP phases 
is  given by\cite{inbtau}
\beqn
\Delta_b^{\tilde g} =-\frac{2\alpha_s}{3\pi}
m_{\tilde g} \tan\beta 
\sum_{i=1,2}\sum_{j=1,2}Re(\mu^{*} e^{-i\xi_3})|D^{*}_{b1i} D_{b2j}|^{2}
f(m_{\tilde g}^2,m_{\tilde b_i}^2,m_{\tilde b_j}^2)\nonumber\\
 +\frac{2\alpha_s}{3\pi}m_{\tilde g}
\sum_{i=1}^2\sum_{j=1}^2 |D_{b2j}|^2|D_{b1j}|^2 Re(m_0A_b e^{-i\xi_3})
f(m_{\tilde g}^2,m_{\tilde b_i}^2,m_{\tilde b_j}^2)
 \eeqn
 where D is defined so that $\tilde b_L=\sum_{i=1}^{2} D_{b1i} \tilde b_i$ and  
$\tilde b_R=\sum_{i=1}^{2} D_{b2i} \tilde b_i$ where 
$\tilde b_i$ are the b squark mass eigen states and  f is defined by
\beq
f(a,b,c)
=\frac{abln(a/b)+bcln(b/c)+acln(c/a)}{(a-b)(b-c)(a-c)}
\eeq 
This result is valid for arbitrary values of $\tan\beta$.
To compare our result to previous analyses we  
set the phases to zero and take the large $\tan\beta$
limit. In this limit we find 
\beq
\Delta_b^{\tilde g}= \frac{2\alpha_3\mu M_{\tilde g}}
{3\pi}   \tan\beta f(m_{\tilde b_1}^2, m_{\tilde b_2}^2,M_{\tilde g}^2)
\eeq
The above result is exactly what is obtained in previous analyses
in the limit of no phases and large  $\tan\beta$\cite{Hall:1993gn,carena94}. 
Similarly the chargino exchange contribution in the limit when 
$\tan\beta$ is  large and the phases are set to zero gives\cite{inbtau}
 
\beq
\Delta_b^{\tilde \chi^+}= \frac{Y_t\mu m_0 A_t}
{4\pi} \tan\beta f(m_{\tilde t_1}^2, m_{\tilde t_2}^2,\mu^2)
\eeq
where $Y_t=h_t^2/4\pi$. Again the above result agrees with the result
of previous analyses without phases valid for large 
$\tan\beta$\cite{Hall:1993gn,carena94}.
One notices that both the limiting forms of the gluino and the chargino 
exchange contributions depend linearly on $\tan\beta$. Because of this
the loop corrections to the $b$ quark mass can become large for 
large $\tan\beta$. A very similar result holds for the case of 
the $\tau$ lepton except that here one only has chargino and neutralino
exchange contributions as exhibited in Fig.~(\ref{tauloop}). 
For the top quark the SUSY QCD and SUSY electroweak corrections have
a very different dependence on $\tan\beta$. Thus the  top mass at 
the Z scale is given by 
\beq 
m_t(M_Z)=\lambda_t(M_Z)\frac{v}{\sqrt 2}\sin\beta(1+\Delta_t) 
\eeq
where $\Delta_t$ gives the loop correction to $m_t$ which arise from
the loop corrections involving the gluino, chargino and neutralino
exchange contributions. The effective $ttH^0$ interaction in this
case is given by 
\beqn
-L_{ttH^0}=(h_t+\delta h_t)\bar t_R t_L H_2^0 
+\Delta h_t \bar t_R t_L H_1^0 + H.c.
\eeqn
where 
\beqn
\Delta_t = (\frac{Re (\Delta h_t)}{h_t} cot\beta +
\frac{Re (\delta h_t)}{h_t} )
\label{deltatop}
\eeqn
The computation for $\Delta_t$ 
 in the presence of CP phases is given in Ref.\cite{inbtau}. 
 Here we observe
 that unlike the $b$ quark case the first term in Eq.~(\ref{deltatop})
 is suppressed because of $cot\beta$ factor rather than enhanced 
 when $\tan\beta$ gets large. Consequently the size of SUSY loop correction to 
 the $t$ quark mass is much smaller than for the case of the
 $b$ quark mass. \\
 
 \begin{center} 
\begin{tabular}{|c|c|c|}
\multicolumn{3}{c}{Table 1. Electron, neutron and $H_g$ edms 
(From Ref.\cite{inbtau})} \\
\hline
\hline
case & $m_0$, $m_{\frac{1}{2}}$, $|A_0|$ & 
$\alpha_A$, $\xi_1$, $\xi_2$, $\xi_3$ \\
\hline
\hline
(a) & $200, 200, 4$ & $1, .5, .659, .633$  \\
\hline 
(b) & $370, 370, 4$ & $2, .6, .653, .672$ \\
\hline
(c) & $320, 320, 3$ & $.8, .4, .668, .6$ \\
\hline 
\hline
 $d_e(ecm)$ & $d_n(ecm)$& $C_{H_g}(cm)$\\
 \hline
(a) ~~$1.45\times 10^{-27}$ 
& $9.2\times  10^{-27}$ &   $7.2\times  10^{-27}$ \\ 
\hline
(b) $-1.14\times 10^{-27}$ &
$-7.9\times 10^{-27}$ & $2.87\times 10^{-26}$ \\
\hline
(c) ~$-3.5\times 10^{-27}$ &
$7.1\times 10^{-27}$ &  $2.9\times 10^{-26}$ \\
\hline
\end{tabular}
\end{center}
\begin{center}
Table Caption:{ $\theta_{\mu}=2.5$ rad for cases (a), (b), (c) in the table.}
\end{center}

\section{Discussion of results}
\begin{figure}
\vspace{-2.5cm}
\hspace*{-0.6in}
\centering
\includegraphics[width=8cm,height=8cm]{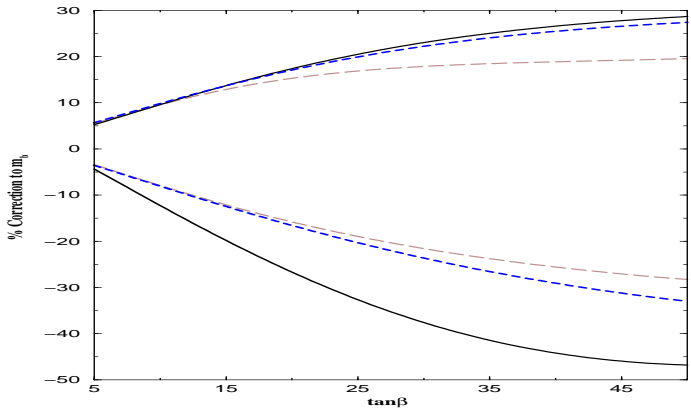}
\caption{Exhibition  of the b quark mass correction  $\Delta m_b/m_b$ 
in percentage  
as a function of $\tan\beta$ with the other parameters given by the
three cases (a), (b) and (c) of Table 1. In the lower half  plane,
the long dashed curve corresponds to the case (a), 
the solid curve corresponds to the case (b) and 
the dashed curve corresponds to the case (c).We note that in each
of the cases (a), (b) and (c) the edm 
constraints for the electron, the neutron and $Hg^{199}$ are satisfied for 
the case $\tan\beta =50$. Similar upper curves have all the same 
parameters as the lower ones except that  the phases are all
set to zero. The figures is taken from Ref.[31]}
\label{bfigE}
\end{figure}

\begin{figure}
\vspace{-2cm}
\hspace*{-0.6in}
\centering
\includegraphics[width=8cm,height=8cm]{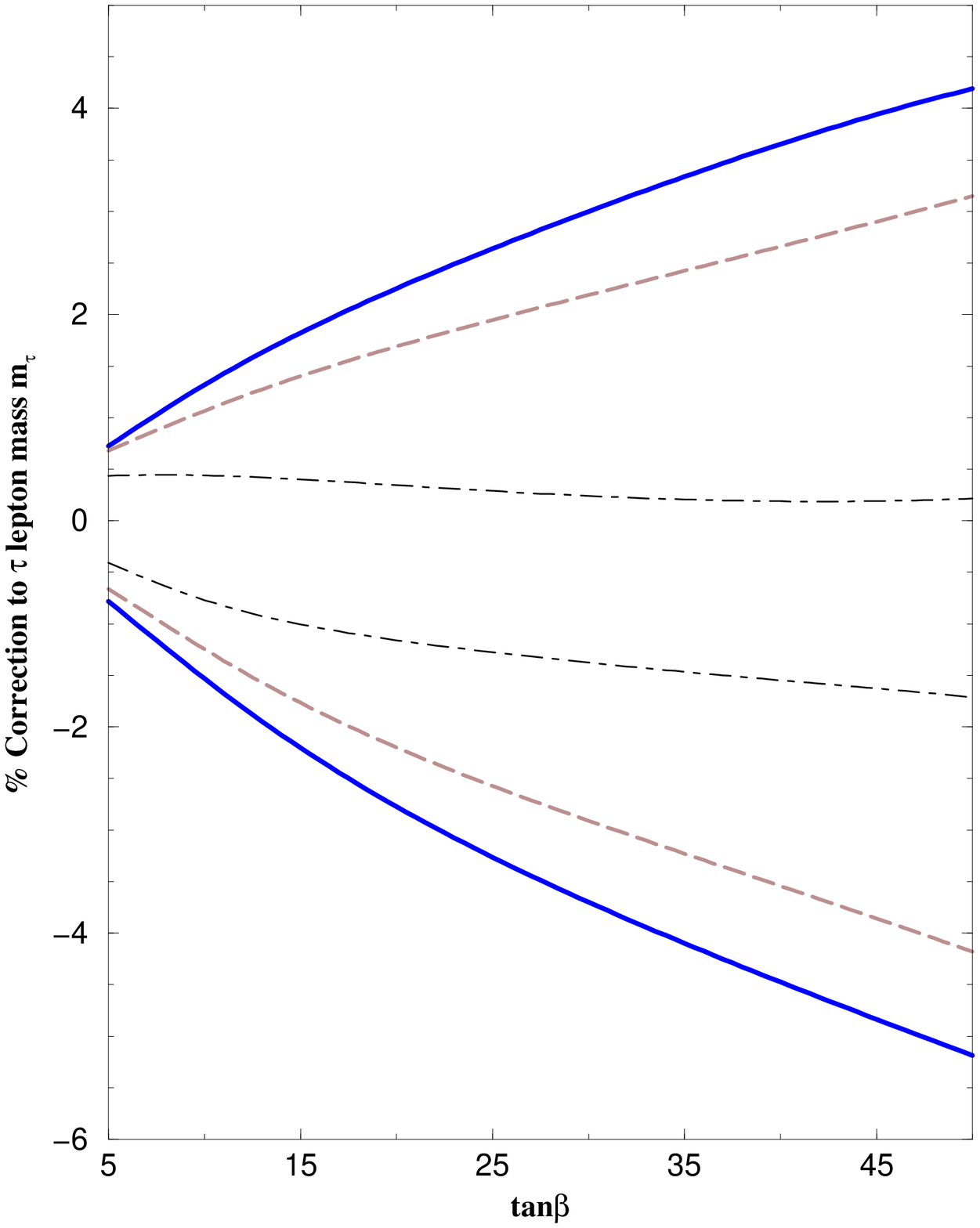}
\caption{Same as  Fig.~(\ref{bfigE}) except that the plot is
for the $\tau$ lepton mass correction $\Delta m_{\tau}/m_{\tau}$ 
in  percentage as a function of $\tan\beta$ for the three
cases (a), (b) and (c) of Table 1.
The figures is taken from Ref.[31]}
\label{taufigE}
\end{figure}

\begin{figure}
\vspace{-2cm}
\hspace*{-0.6in}
\centering
\includegraphics[width=8cm,height=8cm]{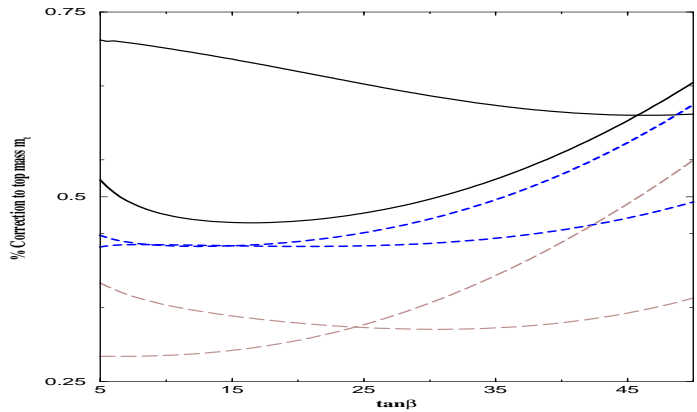}
\caption{Same as  Fig.~(\ref{bfigE}) except that the plot is
for the $t$ quark mass correction $\Delta m_t/m_t$ 
in  percentage as a function of $\tan\beta$ for the three
cases (a), (b) and (c) of Table 1.The figures is taken from Ref.[31]}
\label{tfigE}
\end{figure}

In the numerical analysis we will assume SUGRA models with nonuniversalities 
consistent with the FCNC constraints. Specifically, we assume 
that the parameter space of the model is defined
by the parameters $m_0$, $m_{\frac{1}{2}}$, $A_0$, $\tan\beta$,
$\theta_{\mu}$ and  $\xi_i$ (i=1,2,3). 
We begin by discussing the satisfaction of the edm constraints
which are already rather stringent.
Thus for the electron the current experimental limit on the edm is 
$d_e< 4.3\times 10^{-27}ecm$\cite{eedm} while for the neutron  it is
$ d_n<6.5\times 10^{-26} ~ecm$\cite{nedm}. As noted earlier 
the atomic edm constraints are also now very stringent. Thus for
 $Hg^{199}$  atom  one has $d_{Hg}<9\times 10^{-28} ecm$\cite{atomic}.
This EDM constraint could be translated into a constraint on 
a specific combination of the chromo electric dipole moments of
u, d and s quarks so that $C_{Hg}=|d_{d}^C 
- d_{u}^C -0.012 d_{s}^C|$ is constrained to satisfy $C_{Hg}<3.0 
\times 10^{-26}$cm.
In Table 1 we present cases where in SUGRA models with 
nonuniversalities you have satisfaction of the edm constraints
for the  electron, for the neutron and for the $Hg^{199}$ edm.
The three cases in Table 1 all have large phases  typically order
unity and still one has satisfaction of the edm constraints.
We discuss now the effect of these large  phases on the
analysis of supersymmtric contributions to the b quark.
In Fig.~\ref{bfigE} we give an analysis of the SUSY contribution
to the b quark mass as a function of $\tan\beta$ where the other
parameters correspond to the three cases  given in Table 1. 
The lower curves are with phases while the upper curves are without
phases. Fig.~\ref{bfigE} exhibits several interesting features.
First, one finds that the SUSY correction as expected does indeed
increase essentially linearly with $\tan\beta$ for large $\tan\beta$.
Further, one finds that the effects of phases for each of the three
cases is rather drastic in that the sign as well as the magnitude
of the supersymmetric  correction is affected. We also note 
that with the inclusion of the susy correction the effect can be
as much as 50\% or more. This is a rather large correction showing the
importance of the supersymmtric correction as  well as of the phases.
A similar analysis but for the $\tau$ lepton mass is given in 
 Fig.~\ref{taufigE}. Here also one finds as expected an
 essentially linear dependence on $\tan\beta$  for large 
 $\tan\beta$. Further, one also finds that the
phases affect both the sign as well as the magnitude of the
correction to the $\tau$ lepton mass. However, in this case 
one  finds  that the overall correction is typically order a 
few percent. This  smaller relative  correction is  due essentially
to the fact that in this case one does not  have  a SUSY QCD correction but
only  only a SUSY electroweak correction to the $\tau$ lepton mass.  
In Fig.~\ref{tfigE} we give an analysis of the SUSY QCD and
SUSY electroweak correction to the $t$ quark mass. In this case there
is no $\tan\beta$ enhancement of the SUSY mass correction.
Thus although the CP phases still have a  very significant effect
of the SUSY mass correction, the entire correction in this case
in typically less than a percent or so.

\section{Conclusion}
In this talk we have given a brief overview  of the supersymmetric
corrections to the $b$ and $t$ quark masses and to the $\tau$ lepton
mass including the effects of CP phases  which have been ignored
in previous analyses. It is found that
the effects of these phases on the susy correction to the b quark mass 
can be large enough to change both the sign   and the magnitude of the
correction. Including the effects of the CP phases the supersymmetric
correction can be as much as fifty percent or more of the total mass of the 
b quark mass. 
For the case of the $\tau$ lepton the supersymmetric correction, as 
in the case of the $b$ quark, is proportional to $\tan\beta$ for
large $\tan\beta$ and further, it can change both the sign and the
magnitude when one includes the phases. However, in this case the
overall correction including the phases is typically much smaller
than in the case  of the b quark mass, i.e., only of the order of
a few percent  The smaller correction is in part due to the fact
that in this case the correction is only electroweak. 
Similarly, the supersymmetric correction to the top
quark mass is found to be very sensitive to the phases.
In this case the correction does not have a large enhancement
factor for large $\tan\beta$. Thus, overall  the correction 
is typically less than a percent. As pointed out in Sec.1 the 
supersymmetric corrections to quark and lepton masses plays an
important role in $b-t$ and $b-t-\tau$ unification in the
context of grand unified theory. Thus the phases have important 
implications in such analyses. In addition to affecting 
corrections to the quark and lepton masses, the CP phases also
affect the Higgs vertices involving couplings of the Higgs to the
quarks and the leptons. These modifications lead to important
CP  effects on the decay of the Higgs to $b\bar b$, $\tau\bar \tau$
and $c\bar c$\cite{Ibrahim:2003jm}.   Thus accurate measurement of the branching
ratios of the Higgs to $b\bar b$, $\tau\bar \tau$ and $c\bar c$
in future collider experiments can reveal the presence of 
supersymmetry and of CP phases.

\section{Acknowledgments}
This research was supported in part by NSF grant PHY-0139967.


\begin{thebibliography}{999}

\bibitem{arason}
H. Arason, D.J. Castano, B.E. Kesthelyi, S. Mikaelian,
E.J. Piard, P. Ramond, and B.D. Wright, Phys. Rev. Lett. 
{\bf 67}, 2933(1991);
B.~Ananthanarayan, G.~Lazarides and Q.~Shafi,
Phys.\ Rev.\ D {\bf 44}, 1613 (1991);
V. Barger, M.S. Berger, and P. Ohman, Phys. Lett. {\bf B314},
351(1993); Phys. Rev. {\bf D47}, 1093(1993);
T. Dasgupta, P. Mamales and P. Nath, Phys. Rev. {\bf D52},
5366(1995);
D. Pierce, J. Bagger, K. Matchev and R. Zhang, Nucl. Phys. {\bf B491},
3(1997); H. Baer, H. Diaz, J. Ferrandis and X. Tata, Phys. Rev.
{\bf D61}, 111701(2000);
W. de Boer, M. Huber, A.V. Gladyshev, D.I. Kazakov, 
Eur.\ Phys.\ J.\ C {\bf 20}, 689 (2001).

\bibitem{baer}
H. Baer and J. Ferrandis, Phys. Rev. Lett.{\bf 87}, 211803 (2001);
T.~Blazek, R.~Dermisek and S.~Raby,
Phys.\ Rev.\ Lett.\  {\bf 88}, 111804 (2002)
[arXiv:hep-ph/0107097];
S.~Komine and M.~Yamaguchi,
Phys.\ Rev.\ D {\bf 65}, 075013 (2002)
[arXiv:hep-ph/0110032];
U.~Chattopadhyay and P.~Nath,
Phys.\ Rev.\ D {\bf 65}, 075009 (2002)
[arXiv:hep-ph/0110341];
M.~E.~Gomez, G.~Lazarides and C.~Pallis,
Phys.\ Rev.\ D {\bf 67}, 097701 (2003)
[arXiv:hep-ph/0301064].



\bibitem{Hall:1993gn}
L.~J.~Hall, R.~Rattazzi and U.~Sarid,
Phys.\ Rev.\ D {\bf 50}, 7048 (1994)
[arXiv:hep-ph/9306309].

\bibitem{carena94}
M.~Carena, M.~Olechowski, S.~Pokorski and C.~E.~Wagner,
Nucl.\ Phys.\ B {\bf 426}, 269 (1994)
[arXiv:hep-ph/9402253].

\bibitem{Pierce:1996zz}
D.~M.~Pierce, J.~A.~Bagger, K.~T.~Matchev and R.~j.~Zhang,
Nucl.\ Phys.\ B {\bf 491}, 3 (1997)
[arXiv:hep-ph/9606211].

\bibitem{carena2000}
M.~Carena, D.~Garcia, U.~Nierste and C.~E.~Wagner,
Nucl.\ Phys.\ B {\bf 577}, 88 (2000)
[arXiv:hep-ph/9912516].

\bibitem{carena2002}
M.~Carena and H.~E.~Haber,
arXiv:hep-ph/0208209;

\bibitem{nedm}
P.G. Harris et.al., Phys. Rev. Lett. {\bf 82}, 904(1999).

\bibitem{eedm}
E. Commins, et. al., Phys. Rev. {\bf A50}, 2960(1994).

\bibitem{atomic}
S.~K.~Lamoreaux, J.~P.~Jacobs, B.~R.~Heckel, F.~J.~Raab and E.~N.~Fortson,
Phys.\ Rev.\ Lett.\  {\bf 57}, 3125 (1986).

\bibitem{ellis} 
See, e.g., J. Ellis, S. Ferrara and D.V. Nanopoulos, 
Phys. Lett. {\bf B114}, 231(1982). 

\bibitem{na} 
P. Nath, Phys. Rev. Lett.{\bf 66}, 2565(1991); 
Y. Kizukuri and  N. Oshimo, Phys.Rev.{\bf D46},3025(1992).

\bibitem{bdm2}
K.S. Babu, B. Dutta and R. N. Mohapatra, Phys. Rev. {\bf D61}, 
091701(2000).

 \bibitem{incancel}
T. Ibrahim and P. Nath,
 Phys. Rev. {\bf D57}, 478(1998);
   Phys. Rev. {\bf D58}, 111301(1998);
 T. Falk and K Olive, Phys. Lett. {\bf B 439}, 71(1998);
 M. Brhlik, G.J. Good, and G.L. Kane, Phys. Rev. {\bf D59}, 115004
 (1999); A. Bartl, T. Gajdosik, W. Porod, P. Stockinger, and
 H. Stremnitzer,  Phys. Rev. {\bf 60}, 073003(1999);
 S. Pokorski, J. Rosiek and C.A. Savoy, 
 Nucl.Phys. {\bf B570}, 81(2000);
 E.~Accomando, R.~Arnowitt and B.~Dutta,
Phys.\ Rev.\ D {\bf 61}, 115003 (2000)
[arXiv:hep-ph/9907446].
  U. Chattopadhyay, T. Ibrahim, D.P. Roy, Phys.Rev.D64:013004,2001;
 C.~S.~Huang and W.~Liao,
Phys.\ Rev.\ D {\bf 61}, 116002 (2000);
[arXiv:hep-ph/9908246].
ibid, Phys.\ Rev.\ D {\bf 62}, 016008 (2000);
 A.Bartl, T. Gajdosik, E.Lunghi, A. Masiero, W. Porod,
H. Stremnitzer and O. Vives, hep-ph/0103324.
\noindent
 For analyses in the context string and brane models see,
 M. Brhlik, L. Everett, G. Kane and J. Lykken, Phys. Rev.
 Lett. {\bf 83}, 2124, 1999; Phys. Rev. {\bf D62}, 035005(2000);
  E. Accomando, R. Arnowitt and B. Datta, 
Phys. Rev. {\bf D61},  075010(2000).
T. Ibrahim and P. Nath, Phys. Rev. {\bf D61}, 093004(2000).

\bibitem{inbrane}
T.~Ibrahim and P.~Nath,
Phys.\ Rev.\ D {\bf 61}, 093004 (2000)
[arXiv:hep-ph/9910553].

\bibitem{chang}
D. Chang, W-Y.Keung,and A. Pilaftsis, Phys. Rev. Lett. {\bf 82}, 
900(1999). 

\bibitem{olive} 
 T. Falk, K.A. Olive, M. Prospelov, and R. Roiban, Nucl. Phys. 
 {\bf B560}, 3(1999); V.~D.~Barger, T.~Falk, T.~Han, J.~Jiang, T.~Li 
 and T.~Plehn,
Phys.\ Rev.\ D {\bf 64}, 056007 (2001);
S.Abel, S. Khalil, O.Lebedev, Phys. Rev. Lett. {\bf 86}, 5850(2001)

\bibitem{inhg199}
T.~Ibrahim and P.~Nath,
arXiv:hep-ph/0208142.

\bibitem{msugra}
A.H. Chamseddine, R. Arnowitt and P. Nath, \Journal{\PRL}{49}
{970}{1982}; ~R. Barbieri, S. Ferrara and C.A. Savoy, \Journal{\PLB}
{119}{343}{1982}; ~L. Hall, J. Lykken, and S. Weinberg,
\Journal{\PRD}{27}{2359}{1983}:~ P. Nath, R. Arnowitt and A.H. Chamseddine,
\Journal{\NPB}{227}{121}{1983}.

\bibitem{pilaftsis}
A. Pilaftsis, Phys. Rev. {\bf D58}, 096010; Phys. Lett.{\bf B435}, 
88(1998);
~A. Pilaftsis and C.E.M. Wagner, Nucl. Phys. {\bf B553}, 3(1999);
~D.A. Demir, Phys. Rev. {\bf D60}, 055006(1999);
~S.~Y.~Choi, M.~Drees and J.~S.~Lee,
Phys.\ Lett.\ B {\bf 481}, 57 (2000)
[arXiv:hep-ph/0002287];
~M.~Boz,
Mod.\ Phys.\ Lett.\ A {\bf 17}, 215 (2002)
[arXiv:hep-ph/0008052].

\bibitem{inhiggs}
T. Ibrahim and P. Nath,  
Phys.Rev.D63:035009,2001; hep-ph/0008237; 
T.~Ibrahim,
Phys.\ Rev.\ D {\bf 64}, 035009 (2001)
[arXiv:hep-ph/0102218];
T.~Ibrahim and P.~Nath,
arXiv:hep-ph/0204092.
~S.~W.~Ham, S.~K.~Oh, E.~J.~Yoo, C.~M.~Kim and D.~Son,
arXiv:hep-ph/0205244.

\bibitem{Carena:2001fw}
M.~Carena, J.~R.~Ellis, A.~Pilaftsis and C.~E.~Wagner,
Nucl.\ Phys.\ B {\bf 625}, 345 (2002)
[arXiv:hep-ph/0111245].
;
M.~Carena, J.~Ellis, S.~Mrenna, A.~Pilaftsis and C.~E.~Wagner,
arXiv:hep-ph/0211467.

\bibitem{kane}
S.~Mrenna, G.~L.~Kane and L.~T.~Wang,
Phys.\ Lett.\ B {\bf 483}, 175 (2000)
[arXiv:hep-ph/9910477];
A. Dedes, S. Moretti, Phys.Rev.Lett.84:22-25,2000;
 Nucl.Phys.B576:29-55,2000; S.Y.Choi and J.S. Lee, Phys. Rev.{\bf D61},
 111702(2000).

\bibitem{barger}
V. Barger, Tao Han, Tian-Jun Li, Tilman Plehn,
 Phys.Lett.B475:342-350,2000; ~ V. Barger, T. Falk, T. Han, 
 J. Jiang, T. Li, T. Plehn, hep-ph/0101106;

\bibitem{zerwas}
S.~Y.~Choi, M.~Guchait, J.~Kalinowski and P.~M.~Zerwas,
Phys.\ Lett.\ B {\bf 479}, 235 (2000);
[arXiv:hep-ph/0001175];
~S.~Y.~Choi, A.~Djouadi, H.~K.~Dreiner, J.~Kalinowski and P.~M.~Zerwas,
Eur.\ Phys.\ J.\ C {\bf 7}, 123 (1999)
[arXiv:hep-ph/9806279].

\bibitem{ing2}
T.~Ibrahim and P.~Nath,
Phys.\ Rev.\ D {\bf 62}, 015004 (2000)
[arXiv:hep-ph/9908443]
;
Phys.\ Rev.\ D {\bf 61}, 095008 (2000)
[arXiv:hep-ph/9907555];
T.~Ibrahim, U.~Chattopadhyay and P.~Nath,
Phys.\ Rev.\ D {\bf 64}, 016010 (2001)
[arXiv:hep-ph/0102324].

\bibitem{masiero1}
See, e.g., 
A.~Masiero and H.~Murayama,
Phys.\ Rev.\ Lett.\  {\bf 83}, 907 (1999)
[arXiv:hep-ph/9903363];
~D.~A.~Demir, A.~Masiero and O.~Vives,
Phys.\ Lett.\ B {\bf 479}, 230 (2000)
[arXiv:hep-ph/9911337].

\bibitem{dedes}
A.~Dedes and A.~Pilaftsis,
Phys.\ Rev.\ D {\bf 67}, 015012 (2003)
[arXiv:hep-ph/0209306].

\bibitem{recent}
A.~Bartl, S.~Hesselbach, K.~Hidaka, T.~Kernreiter and W.~Porod,
arXiv:hep-ph/0306281;
T.~F.~Feng, T.~Huang, X.~Q.~Li, X.~M.~Zhang and S.~M.~Zhao,
Phys.\ Rev.\ D {\bf 68}, 016004 (2003)
[arXiv:hep-ph/0305290]; R.~Arnowitt, B.~Dutta and B.~Hu,
arXiv:hep-ph/0307152.

\bibitem{insusy02}
T.~Ibrahim and P.~Nath,
arXiv:hep-ph/0210251; 
arXiv:hep-ph/0207213.

\bibitem{inbtau}
T.~Ibrahim and P.~Nath,
Phys.\ Rev.\ D {\bf 67}, 095003 (2003)
[arXiv:hep-ph/0301110].

\bibitem{Ibrahim:2003jm}
T.~Ibrahim and P.~Nath,
arXiv:hep-ph/0305201.

\end{thebibliography}
\end{document}